\documentstyle[prl,aps,epsfig,floats,twocolumn]{revtex}
\def\r{\vec{r}}
\def\L{{\cal L}}
\def\LR{{\cal LR}}  
\def\GR{{\cal GR}}

\begin{document}
\draft

\twocolumn[\hsize\textwidth\columnwidth\hsize\csname
@twocolumnfalse\endcsname

\title{Angular Structure of Lacunarity,  and The Renormalisation Group}
\author{R.C. Ball$^{1}$, G. Caldarelli$^{2,3}$, A. Flammini$^{1,2}$}
\address{$^1$ Dep. of Physics, University of Warwick, CV4 7AL Coventry, UK.}
\address{$^2$ TCM Group, Cavendish Laboratory, Univeristy of Cambridge CB3 0HE Cambridge, UK.}
\address{$^3$ INFM - Unit\`a di Roma 1 "La Sapienza", P.le A. Moro 2,
00185 - Roma, Italy.}
\maketitle
\date{\today}
\maketitle

\begin{abstract}
We  formulate the angular structure of Lacunarity in fractals,  in terms of  a
symmetry reduction of the three point correlation function.  This provides a
rich probe of universality,  and first measurements yield new
evidence in support of the equivalence between self-avoiding walks and
percolation perimeters in two dimensions.

We argue the Lacunarity reveals much of the
Renormalisation Group in real space.  This is supported by exact
calculations for Random Walks and measured data for percolation clusters and
SAW's. Relationships follow between exponents governing inwards and outwards
propagating perturbations,  and we also find a very general test for the
contribution of long range interactions.

\end{abstract}
]
\narrowtext

In this letter we bring together two outstanding issues in the theory of
fractals, which we believe will also have bearing on critical phenomena more
generally. The first is 'Lacunarity' \cite{M1,GMA1,GMA2}, which as originally
introduced could entail all the discriminations of fractal structure (beyond
dimension) which are evidently visible yet hard to codify in a simple generic way.
The second is the Renormalisation Group \cite{RG} which has been a predominant
influence for three decades in the {\it theory} of critical scaling phenomena, 
fractals included,  yet has never to our knowledge been regarded as a directly {\it measurable}
object in its own right.  Here we will present evidence and arguments that a
particular version of Lacunarity measurement is effectively a measurement of the
(Linearised) Renormalisation Group.

The 'Lacunarity Function' was introduced \cite{BB} as a scaling reduction of
the three point correlation function for a statistical fractal,
  \begin{equation}
\L(R,\theta) = \frac{C_3(\r_1,\r_2,\r_3)}{C_2(\r_1,\r_2)C_2(\r_1,\r_3)  }
  \end{equation}
where $C_3(\r_1,\r_2,\r_3)$ is the conditional probability for points
$\r_2$ and $\r_3$ to be occupied given that $\r_1$ is occupied, 
$C_2(\r_1,\r_2)$ the conditional probability for point $\r_2$ to be
occupied given that $\r_1$ is occupied,  and the Lacunarity Function $\L$
depends only on the angle $\theta$ between the two vectors
$\r_2-\r_1$ and $\r_3-\r_1$ and their length ratio 
$R=r_{13}/r_{12}$.  

The Lacunarity function defined above is (by
construction) independent of absolute lengthscale and is a pure critical-point
scaling object. It is crucial that we are discussing measurements averaged over
an ensemble of fractal objects,  or at least self-averaging over the interior of
one object much larger than than both $r_{12}$ and $r_{13}$.  The dependence of
$\L$ on only a {\it ratio} of lengths follows from the assumption of continuous
scale invariance,   whilst the dependence on only a single internal angle
follows in the more restrictive assumption that the ensemble of fractals in
question has no favoured directions. The whole analysis presupposes the
definition of some suitable measure for the fractal structure in question,  but
provided this is not multifractal \cite{Mmultifractal,HMP} any simple ambiguity
in the measure is expected to cancel from the Lacunarity Function $\L$.

\section{The Angular Structure of Lacunarity}

We introduce here the decomposition of the Lacunarity Function 
over angular harmonics,  for example in two dimensions
\begin{equation}
\L(R,\theta)=\L_0(R)+2\sum_{m=1}^\infty{\L_m(R)\cos{m\theta}}.
\end{equation}
Our physical interpretation of the individual components $\L_m(r'/r)$
is that they measure the correlation between different lengthscales $r$ and $r'$ of
relative fluctuations,  of a particular angular harmonic, in the {\it local} two point
correlation about each mass point.  They can be more precisely calculated as ratios
\begin{equation}
\L_m(r'/r)=M_m(r',r)=
\frac{<S_m(r';\r_1)S_{m}(r;\r_1)^*>}{<S_0(r';\r_1)><S_0(r;\r_1)>} 
\label{eq:lr}
\end{equation}
where $S_m(r;\r_1)$ is the weighted count of occupied points within a
shell of radius $r$ about mass point $\r_1$,  each point weighted by the angular
harmonic of its direction relative to $\r_1$  (i.e. ${\rm e}^{{\rm i} m\theta}$ in two
dimensions),  and the averages are with respect to all occupied mass points
$\r_1$. \cite{cost}

The width of the shells used to compute the averages in (\ref{eq:lr}) drops
out of the result provided it is small enough,  but with finite
datasets some compromise must be made to obtain reasonable signal to noise
ratio. We typically used shells of 7\%  width by radius. In the
case of diagonal elements (only),  that is $r=r'$ equivalent to $R=1$, 
shot noise makes a systematically positive contribution to the reading
for which we have corrected our data below.

Figure 1 shows contour plots of the shell mass correlations $M_m(r',r)$ 
as a function of $ln(r)$
and $ln(r')$ for a large site percolation cluster grown  at criticality.  
The structure
along the diagonal is direct evidence for the  reduction 
$M_m(r',r)=\L_m(R=r'/r)$ and on the basis of such plots we have
averaged data parallel to the diagonal,  within suitable looking
windows,  to obtain the curves for $\L_m(R)$ shown in figure 2.

Figure 3 shows a direct comparison between the corresponding
measurements for the outer perimeter set of such a percolation
cluster with those for a large self-avoiding walk.  The broad
agreement (with no adjustable parameters in any of the
cureves shown) reinforces the claim that these two objects are
in the same Universality Class,  as very strongly suggested by 
Duplantier's caclulation \cite{D} of their identical $f(\alpha)$ spectra for
the exterior harmonic measure.

\section{Connection to the Real Space Renormalisation Group}

The physical idea behind the renormalisation group is that under
critical conditions the structure of a system is propagated through a
continuous cascade of length scales.  This is usually viewed in terms
of the propagation of certain (judiciously chosen) renormalised
parameters which remain invariant at the fixed point.  The
linearisation of the renormalisation group about the fixed point then
corresponds to the transmission of perturbations through the
lengthscale cascade.

Our lacunarity function measures the correlation of structure
fluctuations between different lengthscales and it is natural to
suppose that this correlation arises due to propagation.  If so, 
then our measurements of correlation directly probe the linearised
renormalisation group.  More strictly, we need the covariances rather than the correlations
to probe propagation, to which end it is useful to distinguish from
Lacunarity notation (and rescale) by writing
\begin{equation}
\LR_m(R)=\frac{\L_m(R)-\delta_{m0}}{\L_m(1)-\delta_{m0}}, 
\end{equation}
where we propose to take the liberty of refering to $\LR$ as the {\it Local
Renormalisation Group Function}.

Our proposal is  that if the
local two point correlation function structure at radius $r$ about
point $\r_1$ is perturbed $C_2(r)_{\r_1} \rightarrow C_2(r)(1+
\epsilon_m(r;\r_1) H_m)$,  where $H_m$ is a (normalised) angular
harmonic, then the corresponding response at larger lengthscale $r'$
should follow

\begin{equation}
\epsilon_m(r';\r_1) = \LR_m(r'/r)  \epsilon_m(r;\r_1).
\end{equation}

It is important to note that the Local Renormalisation Group Function
defined above is a 'shattered' response function, in that it gives the
response due to changing the two point correlation at {\it one}
locality.  If we make a global relative perturbation $\epsilon_m(r)$
of the local two point correlation at lengthscale $r$ around every
mass point,  then the  growth of corresponding response defines a 
``Global Renormalisation Group Function'' $\GR$ through

\begin{equation}
\epsilon_m(r') =  \GR_m(r'/r)  \epsilon_m(r). 
\end{equation}
From the number of (occupied) localities of radius $r$ contributing
to perturbation in a larger vicinity of radius $r'$,  we expect the
components of $\GR$ to scale as

\begin{equation}
\GR_m(R) \simeq R^D \LR_m(R)
\label{eq:gr}
\end{equation}
for a simple mass fractal of fractal dimension $D$.  Note that only even $m$ are
of physical consequence here,  as the (global) two point correlation function is
by construction an even function restricted to even perturbations.

An important consequence of equation (\ref{eq:gr}) is that global perturbations
are relevant at large lengthscales if the corresponding Local RG function falls
off more slowly than $R^{-D}$ for large $R$.

The case of simple Random Walks in $d>2$ illustrates  and tests these ideas with exact results.  The
general form of the Lacunarity function $\L(R,\theta)$ was first given
in ref\cite{BB},  but the angular harmonic decomposition
of $\LR$ gives the much simpler form:
\begin{equation}
\LR_0(R)= R^{2-d}; \,\, \LR_m(R)=\frac{1+R^{2-d}}{2}R^{-m},\,\, m>0 .
\end{equation}
This gives  $\GR_0(R) \simeq R^{4-d}$,  $\GR_2(R) \simeq R^0$ and 
$\GR_4(R) \simeq R^{-2}$ for large $R$, which all have physical
interpretation. 

A negative $m=0$ perturbation at lengthscale $r$ means depressing the incidence
of other points at radius $r$ from any given one,  corresponding in polymer
language to turning on some excluded volume.  That the resulting perturbation at
larger lengthscales $r'$ grows as $(r'/r)^{4-d}$ corresponds to the well known
Fixman perturbation expansion of Coil Swelling\cite{Fixman}.  The analogous
positive perturbation leads to coil shrinkage (and ultimately collapse).

An $m=2$ perturbation corresponds to the local structure being stretched in some
directions and shrunk in others,  corresponding for a random walk to anisotropy
in the individual walk steps.  The scale independence of the resulting response
follows from the equivalence of such random walks to affine deformations of
isotropic ones,  which we discuss as a much more general result below.
 
An example of an $m=4$ perturbation would be the intrinsic bias of random walks
made on a square (more generally hypercubic) lattice,  and the decay of the
perturbation reflects the known irrelevance of this in the large lengthscale
limit.  Analogous interpretations apply to higher orders of lattice bias,  for
example $m=6$ and the triangular lattice in two dimensions.

An important issue is whether the propagation of perturbation is
necessarily confined to running upwards in lengthscale.  In systems
governed by an equilibrium ensemble we can equally expect reverse
(inwards) propagation governed by the form of $\LR$.  Thus we are led
naturally to predict that for every global perturbation propating
outwards from $r$ to $r'$ as $(r'/r)^{-\sigma}$ there is propagation
inwards of the same symmetry of perturbation from $r'$ to $r$ as
$(r'/r)^{-\tau}$ where $\tau=\sigma+D$.

The corresponding inwards perturbations of a random walk can be
identified.  The zeroth harmonic corresponds to simple confinement
(and checks out), whilst higher harmonics can then be interpreted in
terms of the influence of a confining shape.  The $m=2$ case,  ellipticity, 
is particularly important because this corresponds to polymer elasticity.

Some of these interpretations can also be checked for less trivial fractals. 
For the Self Avoiding Walk (as for the random walk),  the non-trivial part of
the $m=0$ lacunarity is associated with the walk going from lengthscale $r$ out
to lengthscale $r'$ and coming back again.  This is then governed by the 
probability of an SAW to close a loop, which is known to vary as $N^{-1}\propto
{r'}^{-D}$ where $N$ is the number of steps in the walk (to reach radius $r'$). 
This leads to $\GR_0(R) \simeq R^0$ for large $R$,  corresponding (correctly) to
a change in the excluded volume causing the SAW to dilate uniformly on larger
lengthscales. The inset in figure 3 shows our measured data for $L_0(R)-1\propto
\LR_0(R)$ out to large $R$ and is consistent with $R^{-D}=R^{-4/3}$,  albeit the
data is noisy.  The $m=2$ case for the SAW can be interpreted in terms of the
scaling of elasticity for swollen polymers and leads to also to 
$\LR_2(R)\propto R^{-D}$ at large $R$;  this checks quite well against the main
data in figure 3.


We believe a very general result governs the $m=2$ lacunarity due to its RG
connection,  which explains why in all the fractals considered in this letter
(numerically for the percolation cluster of figure 1,  and analytically for 
random and self-avoiding walks),
\begin{equation}
\GR_2(R)\simeq R^0.
\label{eq:gr2}
\end{equation}
This corresponds to the outwards global m=2 perturbation being marginally
relevant,   and it arises because a $m=2$ perturbation can always be
accommodated by a global strain of the metric of space.  This should of course
fail when long range interactions are important,  sensitive to the metric of
distance,  and therefore equation (\ref{eq:gr2})  serves as a general test of
whether long range interactions play a role in the underlying physics.

\section{Concluding Remarks}

We have presented arguments that the Lacunarity Function $\L$,  or at least its
asymptotes,  can be viewed as response functions determined by
the Renormalisation Group. It is harder to firmly establish just how much of the
Renormalisation Group is thereby revealed.  We can certainly restrict any such
claim to the Linearised Renormalisation Group,  because the Lacunarity Function
can only probe fluctuation correlations about the critical point behaviour:  in
RG terms this means that it can only probe the vicinity of the corresponding
fixed point.  We have restricted our attention to simple mass fractals,  directly
characterised in terms of one (positive definite) scalar field,  the mass
density.  We speculate that $\L$ incorporates all the linearised RG relating to
how this field might be coupled locally and bilinearly to itself,  whereas it
cannot know anything about what happens when other fields are introduced.

A rich range of issues remain to be explored.  We already mentioned that the odd
angular harmonics in the Lacunarity Function lack a simple Response Theory
interpretation because the two point correlation function must be even.  One way
to deal with this is to recognise more explicitly the way the Lacunarity
Function depends only on the shape of the triangle of points it correlates, 
which leads to a parameterisation of $\L$ involving only even harmonics;  all
the odd harmonics are thereby related to the even ones.  However it is more
interesting to consider structures from irreversible growth where correlating
present with future growth is the natural correlation to investigate and odd
harmonics are no longer forbidden in such time resolved correlations.  We anticipate that problems like the
response of Diffusion Limited Aggregation to anisotropy should be analysed in
this way.

In principle our work opens the door to measuring the Renormalisation Group, 
via Lacunarity, directly from real experimental data (as opposed to just computer
simulations).  We recognise,  however,  that this may not be easy.  Our
simulations span some three decades of lengthscale (i.e. cluster radii of order
1000 units),  and even then the match up to expected scaling laws is not
always good.  The best way forward would appear to be not to target the
asymptotic exponents,  but rather to compare absolute lacunarity values between
experiment,  simulation and theory at moderate $R$.  On the theory side,  our
exact Random Walk results are extendable to Markov Chains,  but no other exact
results are known (other than exponents).

Finally,  whilst all the fractals discussed in this letter have close relation
to equilibrium critical phenomena, our present definition of Lacunarity certainly
does not extend trivially to critical phenomena in general.  We believe the
correct analogue of our three point correlation for a mass fractal is a four
point correlation of the corresponding critical distributions,  the extra point
being at infinity.  The natural generalisation of our lacunarity function
to spin models would thus be in terms of a four point correlation function at the
critical point,  and we look forward to exploring this in future work.

\section{Acknowledgements}
This work was supportred by EPSRC,  grant GR/L55346, and
 by the EU under TMR contracts ERBFMBI-CT97-2746 and FMRX-CT98-0183.

\begin{figure}
\centerline{\psfig{file=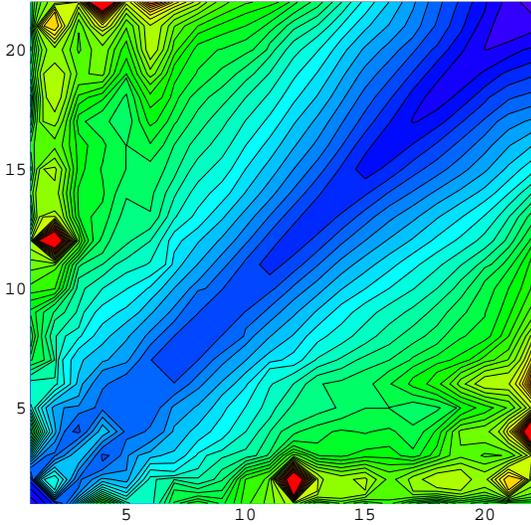,height=7.0cm}}
\caption{Contour plot of the cross correlation of local two point
function between different lengthscales, as per equation (\ref{eq:lr}) with $m=0$,
for a percolation cluster of
$1.5 \,10^6$ sites grown on a traingular lattice at the percolation threshold.  
The axes are
labelled proportional to the logarithm of the two lengthscales correlated.  Contour
lines parallel to the diagonal correspond to this reducing to a scale
invariant 'Lacunarity Function' depending only on the lenghtscale ratio.
Scale invariance breaks down around the edges of the plot where one lengthscale
approaches either the lattice spacing or the sample size. }
\end{figure}
\begin{figure}
\centerline{\psfig{file=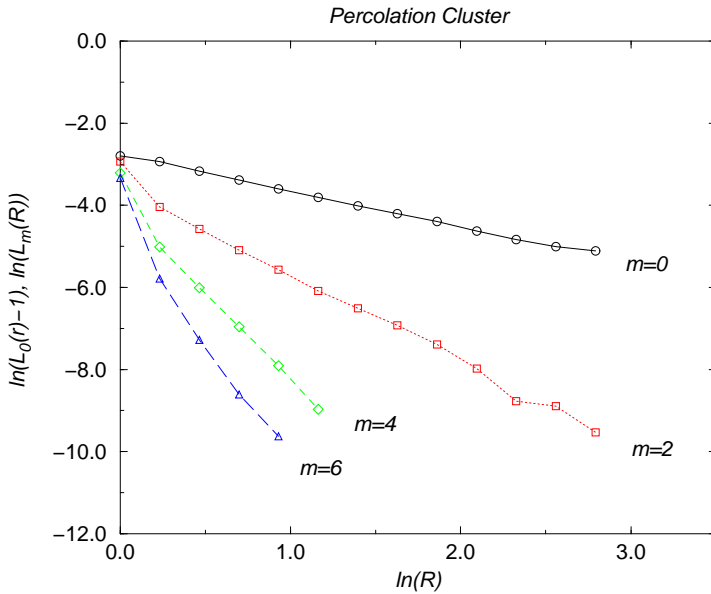,height=8.0cm}}
\caption{The different angular harmonics of the Lacunarity Function
for the two dimensional percolation cluster of figure 1. The  $m=0$ data
correspond to the central part of the contour plot averaged parallel to
the diagonal,  and the other harmonics are defined analogously in
equation (\ref{eq:lr}) }.
\end{figure}

\begin{figure}
\centerline{\psfig{file=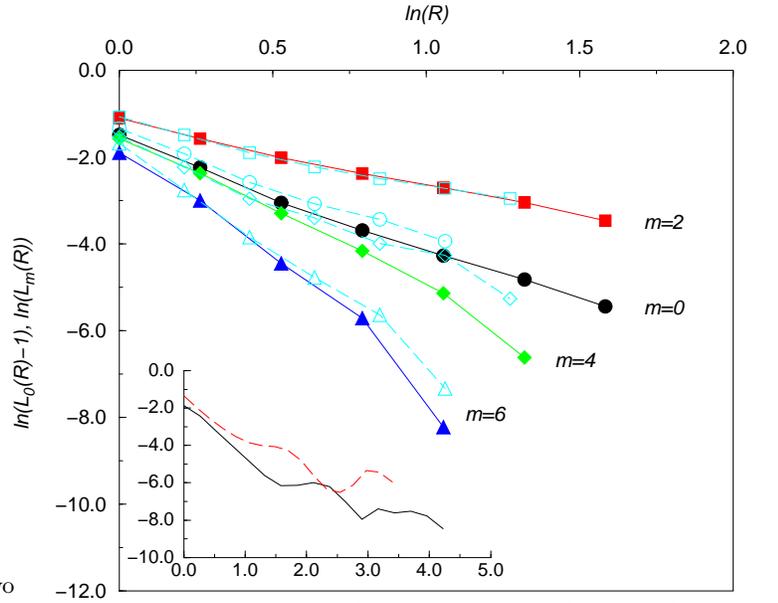,height=8.0cm}}
\caption{Angular harmonics of the Lacunarity Function compared between the
Outer Perimeter of a Percolation Cluster (as per figure 1) and a
Self-Avoiding Walk (in two dimensions) of comparable size.  The
general agreement of these measurements,  which contain no adjustable
parameters, supports both types of object being in the same unversality
class. The inset shows the data for the zeroth harmonic extended to larger lengthscale ratios.}
\end{figure}


\begin{thebibliography}{99}
\bibitem{cost} For a fractal represented by $N$ mass points,  the direct cost of this
operation is Order $N^2$ unless coarse graining economies are used.
\end{thebibliography}
 \end{document}